\documentclass[twocolumn,fleqn]{svjour2}
\usepackage{graphicx}
\usepackage{cite}
\usepackage{bm}         
\usepackage{amsmath}    
\usepackage{amssymb}    
\journalname{Journal of Statistical Physics}

\begin{document}

\title{Rare Events, the Thermodynamic Action and the Continuous-Time Limit}

\author{P.J. Malsom        \and
        F.J. Pinski
}

\institute{P.J. Malsom
           \and
           F.J. Pinski \at
              Department of Physics, University of Cincinnati, Cincinnati, Ohio 45221, USA\\
              Tel.: +1 (513) 556-0523\\
              Fax: +1 (513) 556-3425\\
              \email{frank.pinski@uc.edu}
}


\maketitle

\begin{abstract}
We consider diffusion-like paths that are explored by a particle moving via a conservative force while being in thermal equilibrium with its surroundings.
To probe rare transitions, we use the Onsager-Machlup (OM) functional as a path probability distribution function for double-ended paths that are constrained to start and stop at predesignated points after a fixed time.
We explore the continuous-time limit where the OM functional has been commonly regularized by using the Ito-Girsanov change of measure.
When used as a path measure,  the Ito-Girsanov expression generates an ensemble of double-ended paths that are unphysical.
We expose the underlying reasons why this continuous-time limit does not, and cannot, generate a thermodynamic ensemble of paths. 
Furthermore, we show that the concept of the Most Probable Path and the Thermodynamic action are incompatible with such measures for discrete or continuous time diffusion processes.

\end{abstract}

\section{\label{sec:level0}{Introduction}}

Much of today's work in the study of diffusion processes is grounded in an expression for the probability of a succession of states of a spontaneously fluctuating thermodynamic system.
This expression, famously reported in the 1953 article of Onsager and Machlup{\cite{Onsager:1953}}, has become known as the Onsager-Machlup (OM) functional and has so far withstood the test of time.
One commentary even stated that the results are ``incapable of improvement either in form or in their mode of derivation''{\cite{McKean:98}}.
Here, as in the original article, we represent the fluctuations by white noise, spatially and temporally uncorrelated, and whose amplitude is given by the fluctuation-dissipation theorem.
The underlying motion is then expressed in terms of Brownian dynamics.

For the purposes of this paper, we divide rare events into two groups: those consistent with thermodynamics and those that are not.
As an example of the latter, what we call extremely rare events,  consider the very small probability that all the molecules in a room might migrate to one corner. 
Even though this is allowed by statistical mechanics, if we observed such an event, we would have witnessed a violation of thermodynamics. 
We are not, nor were Onsager and Machlup{\cite{Onsager:1953}}, interested in these extremely rare events.
We are interested in rare, events that are consistent with thermodynamics; ones that are driven by the fluctuations inherent in a thermodynamic system.

In the 1970s, the original OM functional was extended to the continuous-time limit using a Ito's lemma, the Girsanov theorem, and the Radon-Nikodyn derivative. 
The probability measure, which we will refer to as the Ito-Girsanov measure, was written for this limit, and has generated considerable attention{\cite{bach1977functionals,DurrBach:1978,graham1977,
graham1975fluctuations,stratonovich1971probability,horsthemke1975onsager,hunt1981,ito1978probabilistic,
langouche1980short,lavenda1984thermodynamic,tisza1957fluctuations,
watanabe1981onsager,weiss1980uses,yasue1979role}}.
This limiting procedure has been accepted as a method that can be used to look at rare events by constructing probabilities for pinned diffusion paths{\cite{graham1977,dai1989variational,goovaerts2004closed,
ito1984optimal,matsumoto2005exponential,ren2004minimum,watabe1990path}}.
An extension of this limiting procedure, proposed by Graham\cite{graham1977}, and later Eyink\cite{eyink1998action}, stressed the generalization of a ``least-action'' principle to describe particle motions, which then leads to the notion of the Thermodynamic action.

In this paper, we show that this Ito-Girsanov measure, the continuous-time limit of the OM function, is not the appropriate probability measure for exploring thermodynamically driven transitions. 
Furthermore, we show that the probabilities that arise from such ``thermodynamic actions'' are independent of the details of the particle motions as the probabilities are a result of the noise that originates in the thermal bath.
Thus there is no ``action'' to minimize.
We explain how the Ito-Girsanov functional  is unphysical in that it is inconsistent with thermodynamics, and our direct sampling confirms this.
In addition, for thermodynamic significant events, the path-probability measure is flat, that is, it attains the same value for any event that is consistent with thermodynamics.
The point that is missed in the long list of works, cited above, is that the Ito-Girsanov expression only provides the change of measure relative to the free Brownian Bridge. Nowhere has it been proven that it can be used to give the correct relative probability of any two paths. 
On the contrary, here we show that the Ito-Girsanov measure is not a path-probability measure, it is only an indicator of how much a free-Brownian solution differs from a solution for a nonzero force. 

\section{\label{sec:level2}{Brownian Dynamics and the Continuous-Time Path Probability}}

Throughout this paper, we will consider a particle in contact with a heat reservoir at a temperature $\epsilon$.
It is moving under the influence of a potential ${\mathbb{V}}(x)$ with the force being $F(x)= -{\mathbb{V}}'(x)$.
Note that although the equations are written for the one-dimensional case for clarity, the formalism can easily be extended to higher dimensions and for a collection of particles.

The equation of motion for Brownian dynamics is given by the Stochastic Differential Equation (SDE):
\begin{equation}
    dx = F(x) \,dt + \sqrt{2 \, \epsilon} \, dW_t
    \label{SDE}
\end{equation}
 where $dW_t$ is the standard Wiener process that represents the (uncorrelated) Gaussian noise.
Using a discrete time step, $\Delta t$, one typically uses the Euler-Maruyama algorithm{\cite{Maruyama1955}} as an approximate method for propagating the position as a function of time.
In particular,
\begin{equation*}
    x_{i+1} = x_i + F(x_i)\, \Delta t + \sqrt{2 \, \epsilon \, \Delta t \,} \, \xi_i
\end{equation*}
where $\xi_i$ is a Gaussian random variate with mean zero and unit variance.
Successive application (N times) of this equation produces a sequence of positions $\{x_i\}$ which is called a path.
Onsager and Machlup{\cite{Onsager:1953}} used the underlying thermal fluctuations to write the Gaussian path probability, $\mathbb{P}_p \propto \Pi_i \exp{\!( - \frac{1}{2}  \xi_i^2 )} $, in terms of the path variables themselves, namely,
\begin{equation}
    - \ln{ \mathbb{P}_p} ={\mathbb{C}}+
    \frac{\Delta t}{ 2\, \epsilon}\, \sum_{i=1}^N
    \frac{1}{2} \Big( \, \frac{ x_{i+1}-x_i}{\Delta t} - F(x_i)  \Big)^2 \, ,
    \label{ProbOM}
\end{equation}
where ${\mathbb{C}}$ is a constant which is unimportant for this paper.
This equation defines what is sometimes called the thermodynamic action and other times, the OM functional.
In the continuous-time limit, using Ito calculus and the Girsanov theorem, the Radon-Nikodym derivative is used to express the change in the measure\cite{Oksendal2003}.
We define the Ito-Girsanov measure to be
\begin{equation}
    \frac{d \overline{\mathbb{P}}_{p}}{d \mathbb{Q}_{p}} =  \exp{  \!  \Bigg( \! \!-\frac{1}{2 \, \epsilon}
    \Big(  \mathbb{V}(x_T)  - \mathbb{V}(x_0)  + \!  \int_0^T \! \! \!\!  dt \ G(x_t)   \Big) \Bigg)  }
    \label{ProbIto}
\end{equation}
where $T$ is the duration of the path, $\overline{\mathbb{P}}_{p}$ is the continuous-time limit of $\mathbb{P}_{p}$, $\mathbb{Q}_p$ is the measure associated with free Brownian motion, and the function $G(x)$ is defined as  $G(x)=\frac{1}{2} F(x) \cdot F(x)  - \epsilon \,  \mathbb{V}''(x)$.

\section{\label{sec:level88}{Away from the Continuous-Time limit}}

One of the uses of the OM functional is to incorporate it into a scheme to sample paths that are constrained at both ends.
The aim is to efficiently generate an ensemble of paths that include a transition over an energy barrier.
When the barrier is large compared to the typical thermal energy, the transition is a rare event.
We wish to explore barrier hopping that is consistent with thermodynamics, where the driving noise reflects the fluctuating random effects that originate in a thermal bath and thus are independent of the particle's position.

The quandary is that for a very simple one-dimensional example, the generated paths quickly become unphysical when using the Ito-Girsanov form, $\overline{\mathbb{P}}_{p}$, for the path probability.
Long paths, generated with small time steps, are expected to be consistent with equilibrium thermodynamics.
First we address the question of what happens away from the continuous-time limit and show that it is possible to use a method which generates a collection of paths that are consistent with the Boltzmann distribution.
Then we show that when we use Ito-Girsanov expression, given by Eq. $\ref{ProbIto}$, we obtain results that are unphysical.

\begin{figure}[t]
\includegraphics[scale=.6]{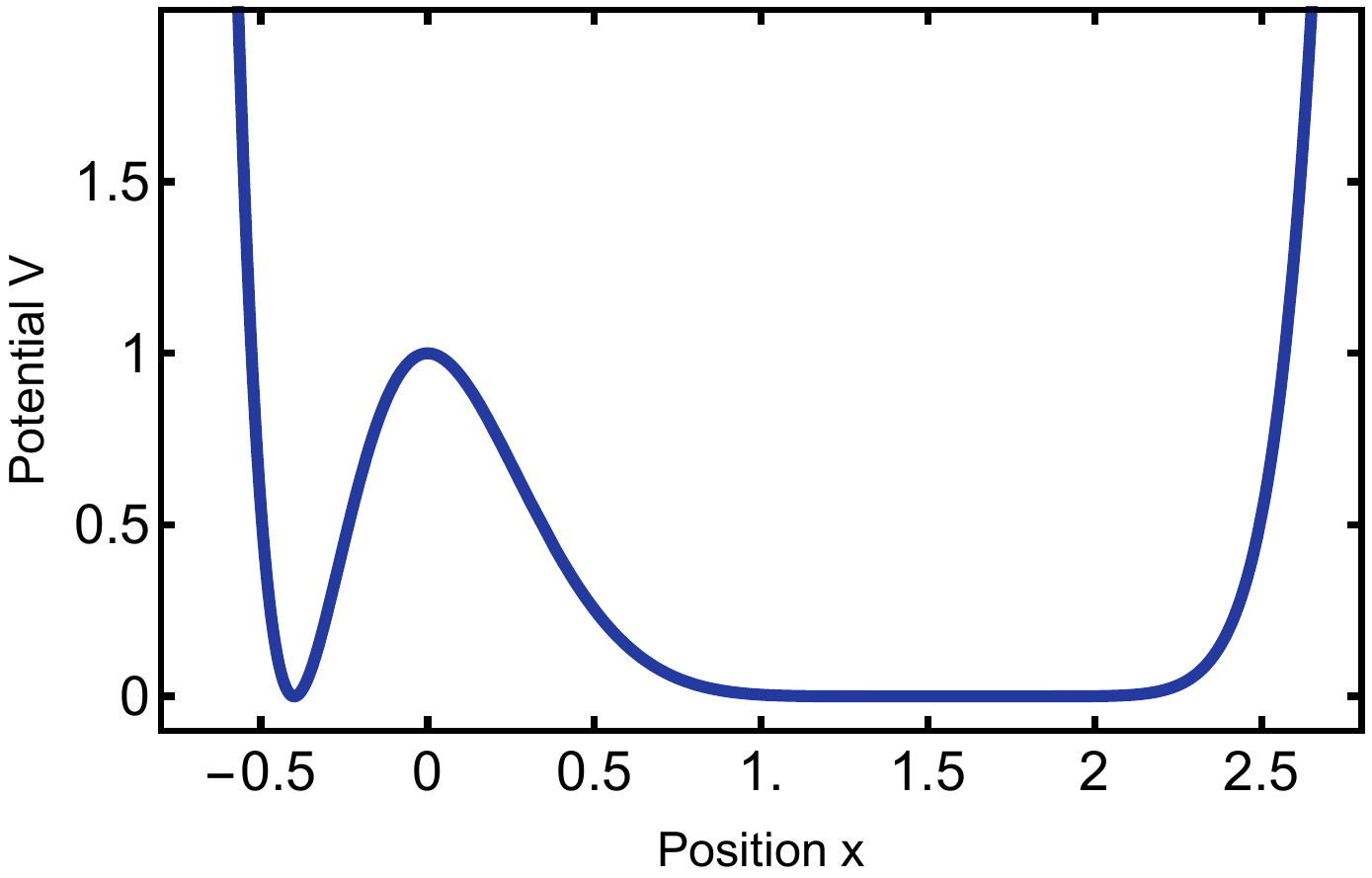}
\caption{
        A plot of the double-well potential used in this paper (see Eq. \ref{ThinBroadPot}). A narrow (quadratic) well, on the left, is separated from the broad well on the right by an unit energy barrier.
            }
    \label{Potential}
\end{figure}

We have constructed a one-dimensional potential to highlight one of the key elements in this problem,
namely, the role of entropy.
We explore whether entropic effects are properly included by forming a potential having two degenerate wells with different widths.
In particular, consider the potential  (plotted in Fig. $\ref{Potential}$)
\begin{equation}
    {\mathbb{V}}(x)  = 2^{-26}  (    8 -5 x   )^8 \, ( 2 + 5x )^2 
    \label{ThinBroadPot}
\end{equation}
which has two degenerate wells with a barrier of unity at the origin.
A narrow (quadratic) well is on the left and the wide well is on the right.
By making the wells degenerate, we eliminate the large (exponential) dependence due to energy differences.
And we accentuate the entropic effects by making one well much wider than the other.

To understand how the time discretization affects the numerical solution to the SDE we used both Eq.  {$\ref{ProbOM}$} and the following form of the path probability $\mathbb{P}_p^s$, 
\begin{equation}
   - \ln{ \mathbb{P}_p^s} =   {\mathbb{C}}  +
    \frac{\Delta t}{ 2\, \epsilon}\, \sum_{i=1}^N  \Bigg(
    \frac{1}{2} \Big|   \frac{ x_{i+1}-x_i}{\Delta t} - F(\overline{x}_i)  \Big|^2 -
     J_i  \Bigg)   
    \label{ProbMP}
\end{equation}
where $J_i=    \frac{2 \, \epsilon}{\Delta t}  \ln \!\Big( 1- \frac{\Delta t}{2} F'(\overline{x}_i) \Big)$,  
 $\overline{x}_i$ is the midpoint of $ x_{i}$ and $ x_{i+1}$.
 This expression,  $\mathbb{P}_p^s$, is related to the Stratonovich representation of the diffusion process (see Elber and Shalloway \cite{elber2000}).
 As noted by Van Kampen {\cite{van1981ito}}, for the problem described by Eq. {$\ref{SDE}$}, the limits of both the Ito and Stratonovich discretizations are equivalent. 	
 In the continuous-time limit $J_i \Rightarrow  \,\epsilon \, F'(\overline{x}_i) $ which provides the Laplacian term in the definition of $G$, giving the same limit as seen in the Ito-Girsanov expression in Eq. {$\ref{ProbIto}$}.
 
In sampling  $\mathbb{P}_p$ and $\mathbb{P}_p^s$, we generated a sequence of paths at a temperature $\epsilon =0.25$, constrained to start in the narrow well and to end in the broad well  and  used a HMC implementation using the ``Implicit Algorithm'' in Beskos \textit{et al.}{\cite{Beskos2008}}. 

We use the Heaviside function $\Theta$ to define the function $B(s)$ to be the fraction of the path that is contained in the broad well, namely,
\begin{equation*}
    B(s) = \frac{1}{T} \int_0^T dt  \ \Theta(x_t^{(s)})  \approx \frac{1}{N}  \sum_i \Theta(x_i^{(s)})
\end{equation*}
where the sampling index is denoted as $s$, and the corresponding path is $\{\,x_i^{(s)}\}.$

\begin{figure}[b]
\includegraphics[scale=.6]{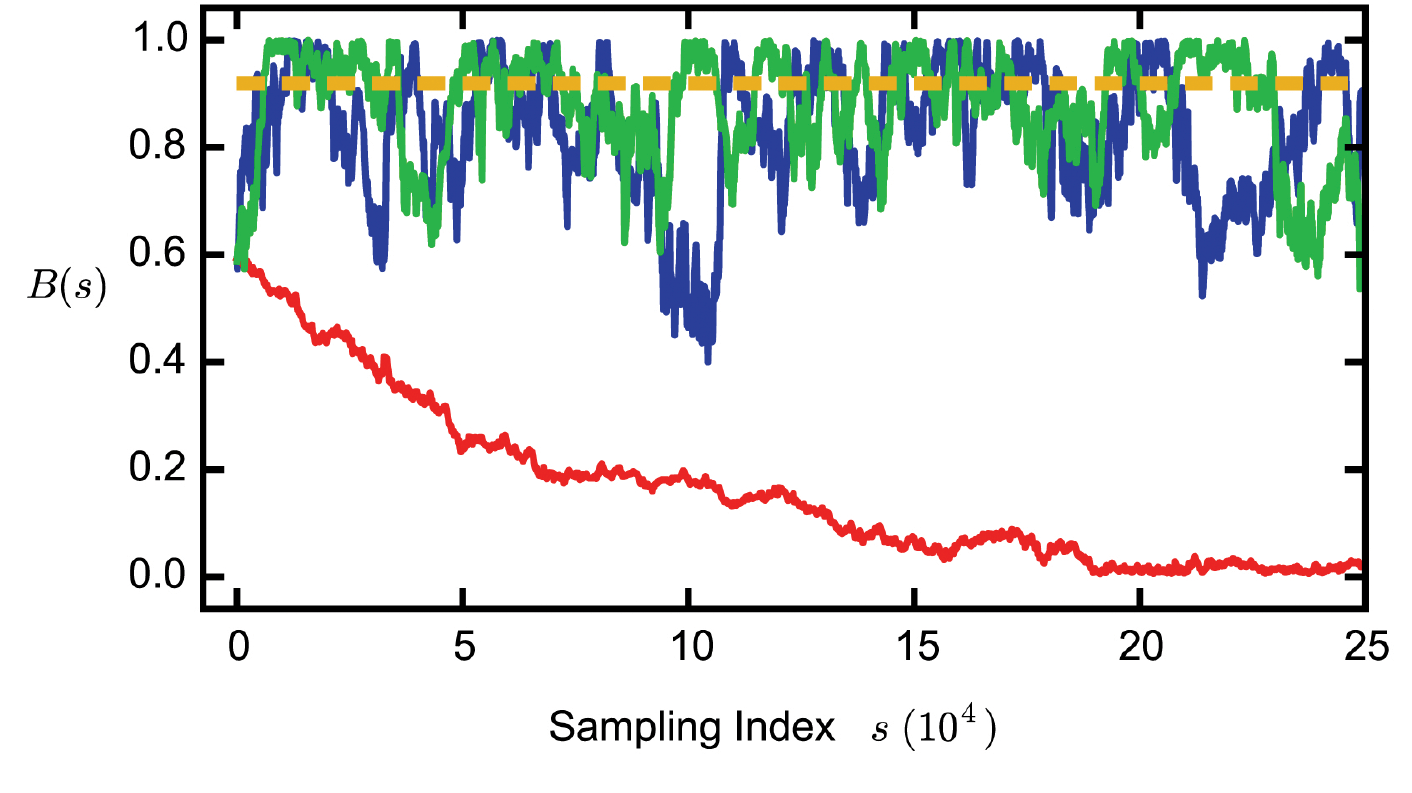}
\caption{
        The fraction of the path in the broad well is shown as a function of an arbitrary sampling index, $s$.
        These are results for three calculation. We started with the same input path where $B(0) \approx 0.6$, and had a path length of $T = N \, \Delta t = 150$, and a time step along the path of $\Delta t = 0.005$.
        The lower (red) curve corresponds to the results using the Ito-Girsanov form, while the upper (green) curve are the results of using the functional (Eq. $\ref{ProbOM}$) 
        with the upper (blue) curve are the results of using the functional (Eq. $\ref{ProbMP}$) based on a Stratonovich  discretization.  
        The (orange) dotted line corresponds to the equilibrium value.
    }
    \label{BroadFractionFig}
\end{figure}

As can be seen in Fig. \ref{BroadFractionFig}, using either of the path probabilities given by a discrete sum, we find an ensemble of paths, which are concentrated in the wide well.
These paths are constrained to undergo at least one transition; we find only one transition in the majority of paths. 
We also find paths which undergo multiple transitions. 
Using either expression, on average, the percent of the time spent in the broad well is approximately $80\%$ which is much closer to the equilibrium value of $90\%$.
This is the type of behavior that one expects for paths consistent with the Boltzmann distribution. 
Away from the continuous-time limit, we found a behavior that is  close to the physical result.
In the next section, we show that these are in stark contrast to calculations that sample the path probability expression given by the Ito-Girsanov expression (Eq. {$\ref{ProbIto}$}).

\section{\label{sec:level2a}{Unphysical Results}}

Now we turn the results found when sampling from the path probability given by  the  Ito-Girsanov expression,  
$\overline{\mathbb{P}}_{p}$ (Eq. {$\ref{ProbIto}$}).   
Results generated using the  Ito-Girsanov expression has been questioned before.
For example in Adib's paper{\cite{adib2008stochastic}}, the origins of the poor performance were not understood.
The potential, described above and used in this work, was designed to highlight entropic effect on transition paths.
With degenerate wells, entropy enhances the time the particle will spent in the wide well.
However, from the form of  $\overline{\mathbb{P}}_{p}$, the curvature of the narrow basin would seem to enhance the probability of paths that spend a long time there, see the early paper of Weiss and H{\"a}ffner {\cite{weiss1980uses}}.
Indeed the Most Probable Path{\cite{MPP:2010}} (MPP) would be highly concentrated in the narrow well.

We used the Ito-Girsanov formula with the method developed by Beskos \textit{et al}{\cite{Beskos}}, referred to here as path-space Hybrid Monte-Carlo (psHMC), using a temperature $\epsilon =0.25$, to generate a sequence of paths constrained to start in the narrow well and to end in the broad well.
In Fig. $\ref{BroadFractionFig}$, we plot $B(s)$ using the described procedure (red curve), showing that the paths quickly become unphysical in that the particle spends the vast majority of the time in the left, thin well.
Such paths are inconsistent with the equilibrium thermodynamical distribution.
Although only one sampling set is presented here, we note that this effect is robust in that similar results are obtained for a range of parameters.

\section{\label{sec:level87}{Random Walk Metropolis}}

\begin{figure}[t]
\includegraphics[scale=.6]{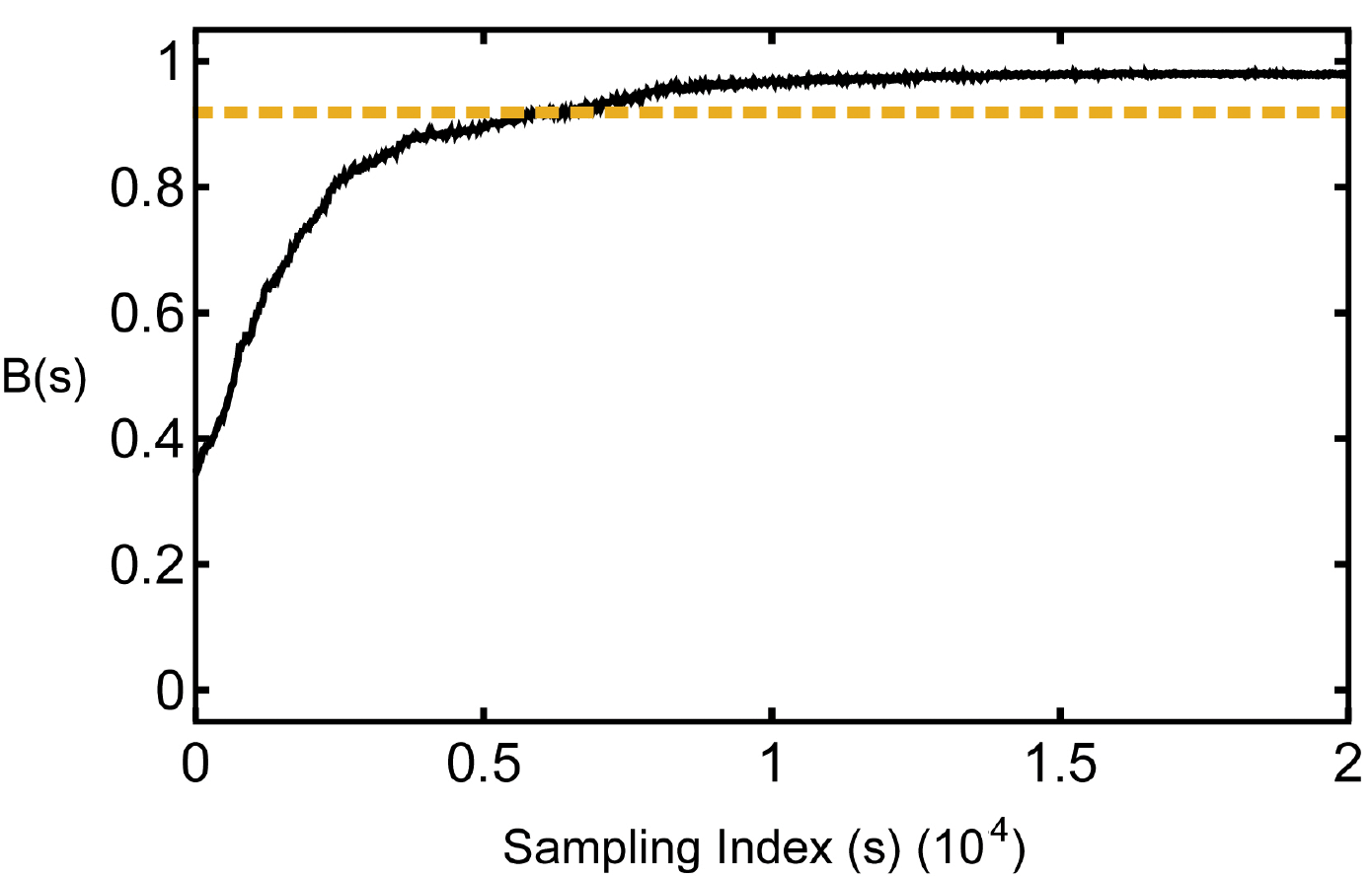}
\caption{
Results for the calculation using the RWM algorithm starting from a randomly generated Brownian Bridge.
As a function of iteration number, a plot of the fraction of the path which resides in the Broad Well (black line) and the equilibrium value (orange dotted line).  }
    \label{RWM}
\end{figure}

We now turn to a simpler sampling method, the Random Walk Metropolis (RWM) algorithm, to sample  $\overline{\mathbb{P}}_{p}$, which proceeds as follows:
First, create a Brownian path that starts and ends at the origin.
A new Brownian Bridge is generated and then combined with the current path (with the sum of the squares of the mixing coefficients being unity).
A proposed path is generated from the Brownian Bridge by shifting the starting point to be $x^-=0.4$ and the ending point to be $x^+=1.6$.
The acceptance or rejection of this new path is based on $\overline{\mathbb{P}}_{p}$ (Eq. $\ref{ProbIto}$). 
This procedure was repeated thousands of times and generated the results displayed in Fig. $\ref{RWM}$.
As shown in this figure, the starting Brownian Bridge evolves to a path that is concentrated in the broad well.
The fraction of the time that the path spends in the positive half plane begins at a value below $0.40$ and settles to a value larger than  $0.95$.
This result differs from the sampling of the Ito-Girsanov expression shown above.
However, if one starts with a path that is highly concentrated ($94\%$) in the narrow well instead of a pinned Brownian path,  one finds that the path does not evolve significantly.  
Looking at Eq. $\ref{ProbIto}$, one finds that the integral of $G$ over the initial path is a large negative number.
The mixing parameter in this case had to be reduced by an order of magnitude otherwise any proposed path was rejected by the Metropolis step.
Clearly, the integral of $G$ derives its large negative value from the path being concentrated in the narrow well.
Looking at the sine transform of the path, one would see that the low frequency part of the spectrum looks more like that of an Ornstein-Uhlenbeck (OU) process{\cite{OU:1930}}  rather than that of free Brownian motion.  
Thus the proposed moves generated in the RWM method are ill-suited for probing the fluctuations around this sharp probability maximum.

\section{\label{sec:level33}{Numerical Aspects}}

To recap the results presented above: we found that by sampling pinned paths we were able to generate different results.
On the one hand, we generated an ensemble of paths that were inconsistent with the Boltzmann distribution by using the Ito-Girsanov expression  and  1) using psHMC algorithm or 2) with the RWM starting from a very particular initial path.
On the other hand, we generated paths that concentrate in the wide well, using a form of the OM functional  with an HMC algorithm 3) with an Ito discretization or 4) with a Stratonovich discretization, away from the continuum limit, and 5) using the Ito-Girsanov expression with the RWM starting from a Brownian Bridge.

The key to understanding numerics behind these results is to examine the ``cross term'' in Eq. $\ref{ProbOM}$. This term can be written as
\begin{align*}
 -   \frac{ x_{ i+1} -x_{ i} }{\Delta t} \cdot F(x_i)  &=  \frac{ x_{ i+1} -x_{ i} }{2 \,\Delta t} \cdot \Big(  F(x_{i+1})- F(x_i)\Big)
 \\  &-  \frac{ x_{ i+1} -x_{ i} }{2 \,\Delta t} \cdot \Big(  F(x_{i+1})+ F(x_i)\Big)  
    \label{cross}
\end{align*}                
The last term above is an approximation to the change-in-potential term in Eq. $\ref{ProbIto}$  (as it telescopes under summation).
The previous term is approximately
\begin{equation}
 \frac{ x_{ i+1} -x_{ i} }{2 \,\Delta t} \cdot \Big(  F(x_{i+1})- F(x_i)\Big) \approx 
 \frac{ \Delta x^2 }{2 \,\Delta t} \cdot  F'( \overline{x})
  \label{cross2}
 \end{equation}
 where $\overline x$ is the midpoint between $x_i$ and $x_{i+1}$.
 A further simplification can be done, but only if the time increment is small and the noise and positions are uncorrelated.
 For sufficiently small $\Delta t$, one can average over several time steps. 
 Doing so, we arrive at an expression that approaches the Ito-Girsanov expression (Eq. $\ref{ProbIto}$), namely,
 \begin{equation} 
 \frac{ \Delta x^2 }{2 \,\Delta t} \cdot  F'( \overline{x}) \rightarrow \,- \epsilon \   {\mathbb{V}}''( \overline{x}).
 \end{equation}
 Remember that the noise, as embodied in $\Delta x^2$, must be uncorrelated to the positions, as required in the underlying SDE (Eq. $\ref{SDE}$) when probing thermodynamically consistent motions. 
It is important to note that this requirement is satisfied (by construction) when the RWM method is used, but only if the correlation is absent in the initial path.
This means that the some of the symptoms of the unphysical probability measure will remain hidden as the RWM method lacks the flexibility to fully sample the measure.

 In the psHMC algorithm the implied noise history and the particle position are intertwined as one searches for the paths that have the larger probabilities as indicated by the Ito-Girsanov expression. 
During the search, correlations are introduced into the low end of the frequency spectrum while the path evolves under the (deterministic) Hamiltonian flow.
High frequencies are correctly handled as paths generated by both methods have quadratic variations that are close to ideal: 
$\sum \, \Delta x^2 \approx 2 \, \epsilon \, T$.
The implication of such low-frequency correlation is that the thermal bath is no longer independent of the system and thus the fluctuations are no longer thermodynamic, and entropy is generated{\cite{Esposito2010Eprod}.

The resulting paths (not shown here) using the psHMC method look like a noisy version of the so-called most probable path{\cite{MPP:2010}}, MPP. 
Recall that the MPPs were generated by finding the probability maximum implied by the Ito-Girsanov expression.  
Thus we conclude that the unphysical nature of the MPPs found previously{\cite{MPP:2010}} is a direct result of using this unphysical measure.
RWM produces paths that are only mixture of Brownian paths which evidently lack the flexibility to probe the regions that the measure indicates are of high probability.

Understanding this cross term (Eq. $\ref{cross2}$) helps to explain why we see less of a problem with calculations for an OU process. 
When $F'$ is a constant, the cross term is averaged correctly if the grid spacing is fine enough so that the Quadratic Variation is close to ideal.  

We note that our findings are not touched by the Ito vs. Stratonovich controversy.
In the problem considered here, the noise and the ``diffusion coefficient'' are spatially homogeneous. 
As pointed out by Van Kampen {\cite{van1981ito}} both discretization schemes lead to the same correct limit.
Looking at Eq. $\ref{ProbMP}$, it is clear that in the limit as $\Delta t \rightarrow 0$, we recover the Ito-Girsanov functional, Eq. $\ref{ProbIto}$, if we regularize the expression.
It is the regularization step that hides an assumption.
When using the Radon-Nikodym derivative, the Ito-Girsanov measure cannot be the probability measure because it has meaning only when using free Brownian Bridges and not for a general path. 

\section{\label{sec:level44}{Equivalence of Path Probability}}

We now turn to a more general discussion of the idea of an MPP in the OM formalism.
We first examine the relative probability of paths.
As in any continuous distribution, while the probability of any one path is zero, the relative probability of two paths is well-defined.
In the case of free Brownian motion, for sufficiently long paths of the same duration, the relative probability of any two paths is unity.
But this is also true when considering Brownian dynamics with any force.
In all cases, the path probability is governed by the Gaussian noise, $\mathbb{P}_p \propto \Pi_i \exp{\!( - \frac{1}{2}  \xi_i^2 )} $,
where the set $\{ \xi_i \}$ is independent of the force and thus independent of the position of the particle.
The same argument applies to any finite representation of the path, and thus holds as the size of the time increment becomes infinitesimally small.

\begin{figure}[t]
\includegraphics[scale=.6]{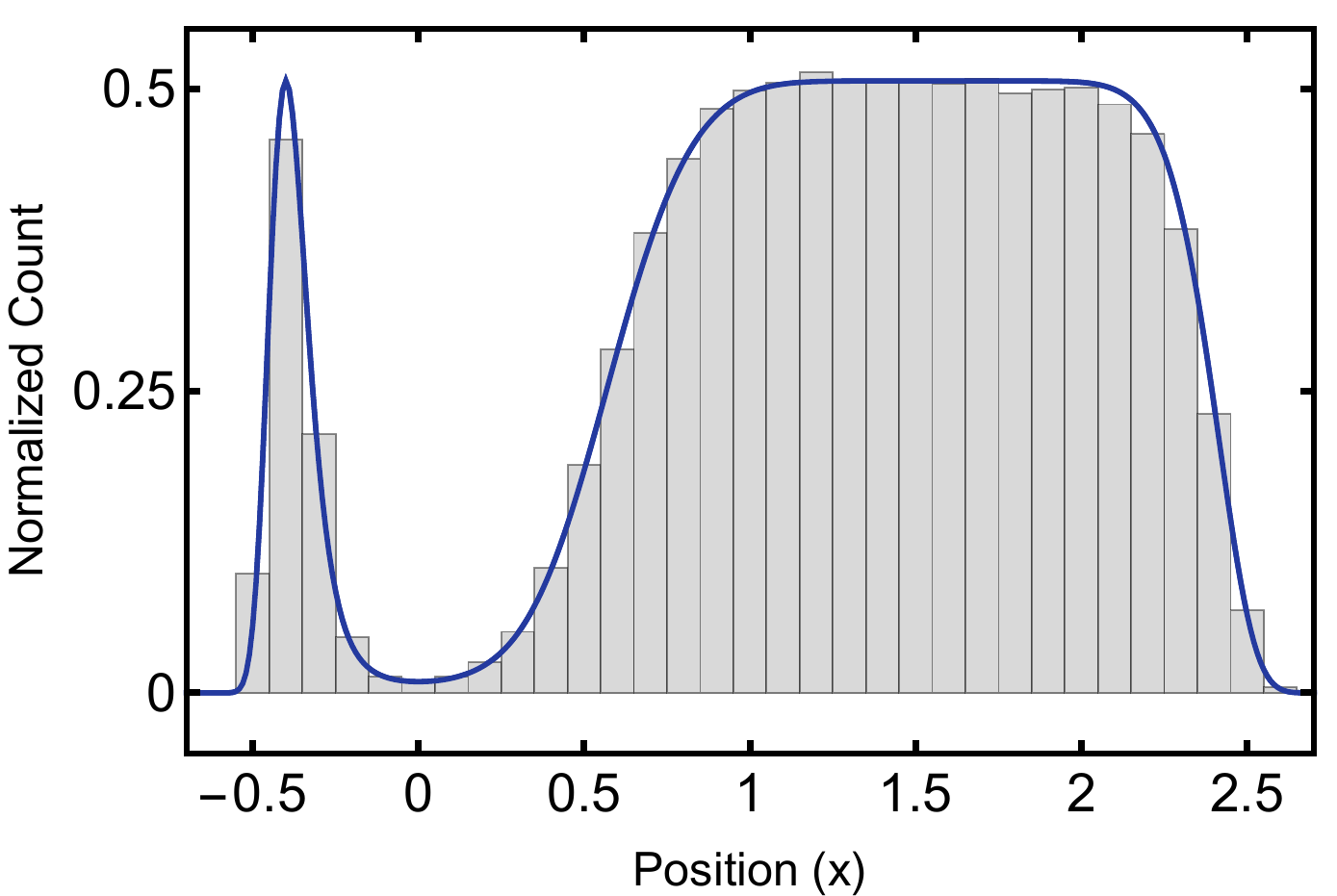}
\caption{
Distribution of endpoints of the set of trajectories with equal  probability.   
Here  the  endpoints  of  472,640  different  trajectories  are  shown.   
The  blue  line  is  the normalized Boltzmann distribution, $\exp{(-\mathbb{V}/ \epsilon)} /Z$, for the temperature $\epsilon = 0.25$, and potential given by Eq. $\ref{ThinBroadPot}$. }
    \label{Histrogram}
\end{figure}

To illustrate the above point, consider the following ``thought'' experiment.
For a conservative force consider sampling the Boltzmann distribution using the Brownian dynamics as expressed by Eq. \ref{SDE}.
Take the starting point, $x_0$, to be arbitrary and integrate the SDE over a fixed time, T, that is long compared to any barrier hopping time.
Using a non-zero but small time step, $\Delta t$, one uses $N_r=T/\Delta t$ Gaussian random variates.
Keeping the same set of random numbers, but simply scrambling the order, redo the integration.
This provides $N_r !$ paths each with identical probabilities.
For large enough T and small enough $\Delta t$, the set of endpoints, $x(T)$ should be distributed in a manner that is close to Boltzmann.
Here it is important to recognize that it is not the path probability that creates the distribution, as all paths are identically probable.
Rather it is the path density that drives the correct distribution of the endpoints.
We have partially preformed this experiment (calculation) for the potential defined in Eq. $\ref{ThinBroadPot}$.
We chose $N_r=2~$million random numbers, Gaussian distributed with mean zero and variance one.
Out of the possible $N_r !$ paths we followed over 400,000 of them using the Euler-Maruyama algorithm, with $\Delta t =0.0005$. 
The endpoints are then collected and the histogram is displayed in  figure $\ref{Histrogram}$.
Evidently the distribution looks to be very close to the Boltzmann distribution in spite of the fact that each path has identically the same probability.
Similar results are expected when one generates independent sets of random numbers instead of using the permutations of the original set.

The noise is a consequence of the random fluctuations of the thermal reservoir, and in the SDE given in Eq. \ref{SDE}, the noise is not correlated with the position of the particle.
These considerations point to the inappropriateness of using the Ito-Girsanov change of measure as the path probability functional for generating thermodynamically consistent paths.
By inspecting Eq. \ref{ProbIto}, we see that it gives different probabilities for different paths: some paths are more probable than others.
This is accomplished by correlating the noise with the positions through the function $G$.
It is clearly incorrect to use the Ito-Girsanov expression in this manner.
And in particular, this is the immediate origin of the numerical results displayed above.

\section{\label{sec:level4}{The Meaning of The Change of Measure}}

The question is how do we understand our results in light of the above information?
To accomplish this, we look at the ramifications for an Ornstein-Uhlenbeck (OU) process{\cite{OU:1930}} with the force being $F_{OU} = - \gamma \, x$.
The Ito-Girsanov change of measure indicates that the probability distributions of two diffusions that differ only in their drift term are mutually absolutely continuous.
Exact solutions are known for both the free Brownian motion and the OU process.
For the OU process, the frequency  ($\nu$) spectrum is finite at the origin with a $1/\nu$  tail.
The frequency spectrum of the free particle motion falls off as $1/\nu$ after diverging at the origin.
For a long enough path length, every realization of the Brownian Bridge has this same spectrum.
Since linear combinations of such realizations are still (free) Brownian Bridges, they too have the same spectrum.
Thus each Brownian Bridge and every linear combination of Brownian Bridges have a frequency spectrum that differs from that of an OU process; they are not solutions to the OU SDE.
This holds true for other forces; in general, in the limit as the path length becomes infinite, free Brownian Bridges are not solutions to the SDE (Eq. \ref{SDE}) with nonzero drift.
This then explains why the RWM sampling method is increasingly sluggish as the path length increases.

With this in mind, we now examine the Kullback-Leibler divergence $D_{KL} (P||Q)$ between the measures for two diffusions $P$ and $Q$ that differ in their drift terms.
If we take $Q$ to represent the measure of free Brownian Bridges then the expectation of the Ito-Girsanov expression can be identified as the Kullback-Leibler divergence since both measures have the same ``normalization'' terms as inferred from the discussion in the previous section. 
This then explains why the Ito-Girsanov change of measure should be viewed as an indicator of the differences between the solutions of two SDEs that differ only in their drift term and is not related to the probability of paths of one or the other diffusions.

\section{\label{sec:level20}{Conclusion}}
The Onsager-Machlup functional is based on Brownian dynamics and provides a way of understanding the double ended path sampling problem{\cite{graham1977}}.
It had been accepted that in the continuous-time limit, the Onsager-Machlup functional could be replaced by the Ito-Girsanov change of measure (Eq. \ref{ProbIto}) as a way of handling the infinities inherent in such a limit.
We have shown here in three ways that this is inappropriate.
First, direct sampling gives unphysical results.
Second, in the long path limit, any linear combination of free Brownian paths cannot be a solution to the OU process nor to diffusions with more complicated forces.
Third, interpreting  the Ito-Girsanov change of measure as a probability distribution favors some paths over others even though paths of the same duration must have the same probability.
We determined that the Ito-Girsanov change of measure is simply an indicator of how inaccurately the free Brownian paths represent solutions to the diffusion process with a nonzero drift.
Thus using the Ito-Girsanov change of measure as a probability distribution is not thermodynamically correct.

The sampling of paths is a complicated numerical procedure interlaced with sophisticated mathematics. 
In such problems, it can be difficult to separate conceptual errors from numerical artifacts. 
It was previously observed that the MPPs{\cite{MPP:2010}} for many potentials seemed to be unphysical, which should have been taken as an indication that something was amiss.
Now we understand why. 
The Ito-Girsanov expression is not a path probability measure and thus the minimizers of the Ito-Girsanov expression are not related to physical paths.

The particular potential, used here, was constructed to highlight entropic effects on transition paths. 
Using it, we were able to pinpoint a common misconception.
We have found similar behaviour with other potentials, such as two quadratic wells with differing widths (connected with a third-order spline).
Also, in two dimensions, we studied diffusion in the potential $V(x,y)= (x^2 + y^8 - 1)^2  \, \exp  ( -2 x^2-3 (y+1)^2 ) \,$, which has an almost degenerate channel. 
There we found unphysical results for a temperature $\epsilon = 0.10$; these results are consistent with the MPP in that the path has a tendency to concentrate near the point ${(0,1)}$  where the channel walls are steepest, rather than in the broad wells.
But such details are unimportant in that they again demonstrate similar behavior as found in the one-dimensional case that we have reported here.
The results for the simple double-well potential shown in this paper clearly illustrate that the Ito-Girsanov expression does not correctly describe entropic effects when used as a basis for sampling paths.
 
The findings in this article affect a wide body of work.
As mentioned above, the theory for entropy production in nonequilibrium thermodynamics, such as the work of Speck et al{\cite{speck2012}}, should be re-examined.
Indeed it would seem that even an encyclopedia entry{\cite{mckane2009stochastic}} needs to be modified.
For probing the folding of proteins, the recent work of Fujisaki, et al{\cite{fujisaki2010}} suffered from using the unphysical form of the Onsager-Machlup functional. 
It is unclear how to interpret their results.
Other works{\cite{facc2006,eric2004,hartmann2012efficient,zhang2013importance}} have to be reevaluated in light of these new results.

The results reported here also have ramifications for sampling general measures in infinite dimensions.
The samples or paths in infinite dimension must be constructed to probe the high probability regions as defined by the measure.
Algorithms based on Brownian Bridges have a very limited flexibility in that the noise history is uncorrelated by construction.
Only in HMC methods, high probability regions are explored due to the Hamiltonian flow which allows the high probability, highly correlated, modes to be explored.
The limited sophistication of the sampling algorithms employed in the past has hid some of the results exposed in this article.

The path integrals that appear in diffusion look similar to those that appear in Quantum Mechanics. 
The natural question is what is impact of the current work on Feynman Path Integrals{\cite{feynman1965quantum}}.
The short answer is none.
Firstly the paths in Quantum Mechanics are smooth, while Brownian paths are almost nowhere differentiable. 
Secondly diffusion is a local process while solutions to the Schr{\"o}dinger equation are not.
The time evolution of the wave function is most easily found by using an eigenfunction expansion; the eigenfunctions are dependent on the potential everywhere in space. Diffusion is described by a Markovian process and white noise; local in space and time.
Thirdly, in the path integral formulation of Quantum Mechanics, the path distribution is peaked around the classical path, a smooth curve that is defined by the classical action. 
As we discussed above, each physical (thermodynamically-consistent) diffusion bridge has the same probability.
The most important observation in this regard is that the field-theoretic methods, such as in the review by Smith{\cite{smith2011}}, developed for Feynman integrals are not transferable to classical diffusion.

\begin{acknowledgements}
We wish to especially thank Robin Ball, Andrew Stuart, Hendrik Weber, Gideon Simpson, and Florian Theil for many lengthy conversations. We acknowledge the use of computing resources provided by the Open Science Grid.
\end{acknowledgements}

\bibliographystyle{spmpsci}
\bibliography{MalsomPinski}

\begin{thebibliography}{10}
\providecommand{\url}[1]{{#1}}
\providecommand{\urlprefix}{URL }
\expandafter\ifx\csname urlstyle\endcsname\relax
  \providecommand{\doi}[1]{DOI~\discretionary{}{}{}#1}\else
  \providecommand{\doi}{DOI~\discretionary{}{}{}\begingroup
  \urlstyle{rm}\Url}\fi

\bibitem{adib2008stochastic}
Adib, A.B.: Stochastic actions for diffusive dynamics: Reweighting, sampling,
  and minimization.
\newblock The Journal of Physical Chemistry B \textbf{112}(19), 5910--5916
  (2008)

\bibitem{bach1977functionals}
Bach, A., D{\"u}rr, D., Stawicki, B.: Functionals of paths of a diffusion
  process and the onsager-machlup function.
\newblock Zeitschrift f{\"u}r Physik B Condensed Matter \textbf{26}(2),
  191--193 (1977)

\bibitem{Beskos}
Beskos, A., Pinski, F., Sanz-Serna, J., Stuart, A.: Hybrid monte carlo on
  hilbert spaces.
\newblock Stochastic Processes and their Applications \textbf{121}(10),
  2201--2230 (2011)

\bibitem{Beskos2008}
Beskos, A., Roberts, G., Stuart, A., Voss, J.: Mcmc methods for diffusion
  bridges.
\newblock Stochastics and Dynamics \textbf{08}(03), 319--350 (2008)

\bibitem{dai1989variational}
Dai~Pra, P., Pavon, M.: Variational path-integral representations for the
  density of a diffusion process.
\newblock Stochastics: An International Journal of Probability and Stochastic
  Processes \textbf{26}(4), 205--226 (1989)

\bibitem{DurrBach:1978}
D{\"{u}}rr, D., Bach, A.: The {O}nsager-{M}achlup function as {L}agrangian for
  the most probable path of a diffusion process.
\newblock Comm. Math. Phys. \textbf{60}(2), 153 --170 (1978)

\bibitem{eric2004}
E, W., Ren, W., Vanden-Eijnden, E.: Minimum action method for the study of rare
  events.
\newblock Communications on Pure and Applied Mathematics \textbf{57}(5),
  637--656 (2004).
\newblock \doi{10.1002/cpa.20005}

\bibitem{elber2000}
Elber, R., Shalloway, D.: Temperature dependent reaction coordinates.
\newblock The Journal of Chemical Physics \textbf{112}(13), 5539--5545 (2000).
\newblock \doi{http://dx.doi.org/10.1063/1.481131}

\bibitem{Esposito2010Eprod}
Esposito, M., Lindenberg, K., den Broeck, C.V.: Entropy production as
  correlation between system and reservoir.
\newblock New Journal of Physics \textbf{12}(1), 013,013 (2010).
\newblock \urlprefix\url{http://stacks.iop.org/1367-2630/12/i=1/a=013013}

\bibitem{eyink1998action}
Eyink, G.L.: Action principle in statistical dynamics.
\newblock Progress of Theoretical Physics Supplement \textbf{130}, 77--86
  (1998)

\bibitem{facc2006}
Faccioli, P., Sega, M., Pederiva, F., Orland, H.: Dominant pathways in protein
  folding.
\newblock Phys. Rev. Lett. \textbf{97}, 108,101 (2006)

\bibitem{feynman1965quantum}
Feynman, R., Hibbs, A.: Quantum mechanics and path integrals.
\newblock International series in pure and applied physics. McGraw-Hill (1965).
\newblock \urlprefix\url{https://books.google.com/books?id=14ApAQAAMAAJ}

\bibitem{fujisaki2010}
Fujisaki, H., Shiga, M., Kidera, A.: Onsager-\uppercase{M}achlup action-based
  path sampling and its combination with replica exchange for diffusive and
  multiple pathways.
\newblock The Journal of Chemical Physics \textbf{132}(13), 134101 (2010).
\newblock \doi{http://dx.doi.org/10.1063/1.3372802}

\bibitem{goovaerts2004closed}
Goovaerts, M., De~Schepper, A., Decamps, M.: Closed-form approximations for
  diffusion densities: a path integral approach.
\newblock Journal of computational and applied mathematics \textbf{164},
  337--364 (2004)

\bibitem{graham1977}
Graham, R.: Path integral formulation of general diffusion processes.
\newblock Zeitschrift fur Physik B Condensed Matter \textbf{26}(3), 281--290
  (1977)

\bibitem{graham1975fluctuations}
Graham, R., Riste, T.: Fluctuations, instabilities and phase transitions.
\newblock Plenum, New York p. 215 (1975)

\bibitem{hartmann2012efficient}
Hartmann, C., Sch{\"u}tte, C.: Efficient rare event simulation by optimal
  nonequilibrium forcing.
\newblock Journal of Statistical Mechanics: Theory and Experiment
  \textbf{2012}(11), P11,004 (2012)

\bibitem{horsthemke1975onsager}
Horsthemke, W., Bach, A.: Onsager-machlup function for one dimensional
  nonlinear diffusion processes.
\newblock Zeitschrift f{\"u}r Physik B Condensed Matter \textbf{22}(2),
  189--192 (1975)

\bibitem{hunt1981}
Hunt, K.L.C., Ross, J.: Path integral solutions of stochastic equations for
  nonlinear irreversible processes: The uniqueness of the thermodynamic
  lagrangian.
\newblock The Journal of Chemical Physics \textbf{75}(2), 976--984 (1981).
\newblock \doi{http://dx.doi.org/10.1063/1.442098}

\bibitem{watanabe1981onsager}
Ikeda, N., Watanabe, S.: The {O}nsager-{M}achlup functions for diffusion
  processes.
\newblock In: K.~Matthes (ed.) Stochastic Differential Equations and Diffusion
  Processes, pp. 510--510. WILEY-VCH Verlag (1986).
\newblock \doi{10.1002/bimj.4710280425}.
\newblock \urlprefix\url{http://dx.doi.org/10.1002/bimj.4710280425}

\bibitem{ito1978probabilistic}
Ito, H.: Probabilistic construction of lagrangian of diffusion process and its
  application.
\newblock Progress of Theoretical Physics \textbf{59}(3), 725--741 (1978)

\bibitem{ito1984optimal}
Ito, H.: Optimal gaussian solutions of nonlinear stochastic partial
  differential equations.
\newblock Journal of statistical physics \textbf{37}(5-6), 653--671 (1984)

\bibitem{langouche1980short}
Langouche, F., Roekaerts, D., Tirapegui, E.: Short derivation of feynman
  lagrangian for general diffusion processes.
\newblock Journal of Physics A: Mathematical and General \textbf{13}(2), 449
  (1980)

\bibitem{lavenda1984thermodynamic}
Lavenda, B.: Thermodynamic criteria governing the stability of fluctuating
  paths in the limit of small thermal fluctuations: critical paths in the limit
  of small thermal fluctuations: critical paths and temporal bifurcations.
\newblock Journal of Physics A: Mathematical and General \textbf{17}(17), 3353
  (1984)

\bibitem{Maruyama1955}
Maruyama, G.: Continuous markov processes and stochastic equations.
\newblock Rendiconti del Circolo Matematico di Palermo \textbf{4}(1), 48--90
  (1955).
\newblock \doi{10.1007/BF02846028}

\bibitem{matsumoto2005exponential}
Matsumoto, H., Yor, M., et~al.: Exponential functionals of brownian motion, ii:
  Some related diffusion processes.
\newblock Probability Surveys \textbf{2}, 348--384 (2005)

\bibitem{mckane2009stochastic}
McKane, A.J.: Stochastic processes.
\newblock In: Encyclopedia of Complexity and Systems Science, pp. 8766--8783.
  Springer (2009)

\bibitem{McKean:98}
McKean, H.P.: In: The Collected Works of Lars Onsager: With Commentary, Lars
  Onsager and P.C. Hemmer, pp. 769--771. World Scientific, Singapore (1998)

\bibitem{Oksendal2003}
{\O}ksendal, B.: Stochastic Differential Equations.
\newblock Universitext. Springer Berlin Heidelberg, Berlin, Heidelberg (2003).
\newblock \urlprefix\url{http://link.springer.com/10.1007/978-3-642-14394-6}

\bibitem{Onsager:1953}
Onsager, L., Machlup, S.: Fluctuations and irreversible processes.
\newblock Phys. Rev. \textbf{91}(6), 1505 (1953)

\bibitem{MPP:2010}
Pinski, F.J., Stuart, A.M.: Transition paths in molecules at finite
  temperature.
\newblock The Journal of Chemical Physics \textbf{132}(18), 184104 (2010).
\newblock \doi{10.1063/1.3391160}.
\newblock \urlprefix\url{http://link.aip.org/link/?JCP/132/184104/1}

\bibitem{ren2004minimum}
Ren, W., Vanden-Eijnden, E., et~al.: Minimum action method for the study of
  rare events.
\newblock Communications on pure and applied mathematics \textbf{57}(5),
  637--656 (2004)

\bibitem{smith2011}
Smith, E.: Large-deviation principles, stochastic effective actions, path
  entropies, and the structure and meaning of thermodynamic descriptions.
\newblock Reports on Progress in Physics \textbf{74}(4), 046,601 (2011).
\newblock \urlprefix\url{http://stacks.iop.org/0034-4885/74/i=4/a=046601}

\bibitem{speck2012}
Speck, T., Engel, A., Seifert, U.: The large deviation function for entropy
  production: the optimal trajectory and the role of fluctuations.
\newblock Journal of Statistical Mechanics: Theory and Experiment
  \textbf{2012}(12), P12,001 (2012)

\bibitem{stratonovich1971probability}
Stratonovich, R.: On the probability functional of diffusion processes.
\newblock Selected Trans. in Math. Stat. Prob \textbf{10}, 273--286 (1971)

\bibitem{tisza1957fluctuations}
Tisza, L., Manning, I.: Fluctuations and irreversible thermodynamics.
\newblock Physical Review \textbf{105}(6), 1695 (1957)

\bibitem{OU:1930}
Uhlenbeck, G.E., Ornstein, L.S.: On the theory of the {B}rownian motion.
\newblock Phys. Rev. \textbf{36}, 823--841 (1930).
\newblock \doi{10.1103/PhysRev.36.823}

\bibitem{van1981ito}
Van~Kampen, N.: It{\^o} versus stratonovich.
\newblock Journal of Statistical Physics \textbf{24}(1), 175--187 (1981)

\bibitem{watabe1990path}
Watabe, M., Shibata, F.: Path integral and brownian motion.
\newblock Journal of the Physical Society of Japan \textbf{59}(6), 1905--1908
  (1990)

\bibitem{weiss1980uses}
Weiss, U., H{\"a}ffner, W.: The uses of instantons for diffusion in bistable
  potentials.
\newblock Functional Integration: Theory and Applications p. 311 (1980)

\bibitem{yasue1979role}
Yasue, K.: The role of the onsager--machlup lagrangian in the theory of
  stationary diffusion process.
\newblock Journal of Mathematical Physics \textbf{20}(9), 1861--1864 (1979)

\bibitem{zhang2013importance}
Zhang, W., Hartmann, C., Weber, M., Sch{\"u}tte, C.: Importance sampling in
  path space for diffusion processes.
\newblock Multiscale Model. Sim  (2013)

\end{thebibliography}

\end{document}